\documentstyle[prl,aps,multicol]{revtex}

\begin{document}

\title{Comment on the paper \\
``Decoherence and the theory of continuous quantum measurements'', \\
by M. B. Mensky, Phys. Usp. 41, 923 (1998), arXiv:quant-ph/9812017 v2}
\author{Roberto Onofrio$^\dag$ and Carlo Presilla$^\ddag$}
\address{$^\dag$Dipartimento di Fisica ``Galileo Galilei'', 
Universit\`a di Padova, Via Marzolo 8, 35131 Padova, Italy}
\address{$^\ddag$Dipartimento di Fisica, 
Universit\`a di Roma ``La Sapienza'',
Piazzale A. Moro 2, 00185 Roma, Italy}
\maketitle
\begin{multicols}{2} \narrowtext

Quantum measurements have attracted a renewed attention in the last 
decades due to the improvements in the sensitivity of various 
devices operating close to their quantum limit.
A formalism to deal with continuous quantum measurements, 
the restricted path integral (RPI) approach, has been developed by 
Mensky \cite{M0} and applied, among the others, by us and 
collaborators to various phenomenological contexts, for instance 
the quantum Zeno effect in atomic spectroscopy \cite{OPT,TPO,POT}.

In a recent paper \cite{M1} Mensky reviewed our application 
of the RPI approach to energy measurements. 
He wrote: 
{\em The approach based on RPI and complex Hamiltonians was 
first applied to the measurement of energy in a two-level system in 
Refs. [79,80]} (Refs. \cite{OPT,TPO} of this Comment). 
{\em It was demonstrated that if the energy is measured with a high 
enough accuracy, then the system becomes frozen, and the transitions 
between the levels are no longer possible (the Zeno effect). 
The alternative regimes of measurements have not been 
(and could not have been) studied in these works because of a serious 
methodological error. 
The authors assumed that the result of a continuous measurement is 
expressed by the function $E(t)$, which does not change and coincides 
with one of the energy levels of the system}.
{\em \ldots This error was corrected in Ref. [20]} 
(Ref. \cite{AM} of this Comment),
{\em which has made it possible to carry out a detailed analysis
of a moderately accurate continuous quantum measurement of the
energy, and propose an entirely new type of measurement - monitoring
of a quantum transition}.

In this Comment we point out that Mensky's assessment of our results 
is incorrect. Our claim is simply based on the actual reading of the 
papers quoted by Mensky himself \cite{OPT,TPO}, as well as on 
commenting results from a  paper surprisingly ignored by him in this 
context \cite{POT}.

In Ref. \cite{OPT} at page 136 (bottom) we explicitly wrote:  
{\em  \dots the measurement result. 
This last is not necessarily an eigenvalue $E_n$ of the unmeasured 
system due to the classical uncertainty of the meter (for a detailed 
discussion of the classical properties of the meter see ref. [14])}. 

In Ref. \cite{TPO} from Eq. (1) to Eq. (15) the measurement result 
is represented by a generic time-dependent function $E(t)$.
In order to obtain a simple solution of the system of differential 
equations (15), we then assume $E$ constant but with an arbitrary value.
Only at the end of Section II we consider the special case, $E=E_n$, 
of a measurement result coincident with a system eigenvalue. 
Analogous considerations hold for the two-level system analyzed 
in Section IV of \cite{TPO}. The soundness of our approach is 
(perhaps unintentionally) confirmed by Mensky himself who, 
in \cite{M1}, later rewrites exactly the equations in \cite{TPO}  
(compare Eq. 69 onward in Section 4.1 of \cite{M1} 
to Eq. 1 onward of Section II of \cite{TPO}). 

Some related considerations on the difference between {\it a posteriori} and 
{\it a priori} selective measurements are also in order. 
In Refs. \cite{OPT,TPO} we considered the case of {\it a posteriori}
selective measurements, i.e. measurement processes analyzed at
the end of the measurement when a particular result $E(t)$ 
has been realized (selected).
A probability $P[E]$ associated to each measurement
result $E(t)$ has to be taken into account  
to describe a selective measurement {\it a priori} during its evolution.
This is what Audretsch and Mensky did in their paper \cite{AM}
by considering a probability-dependent RPI. 
However, a detailed analysis of the differences between 
{\it a priori} and {\it a posteriori} selective measurements
had been previously given in \cite{POT}.
In the {\it a posteriori} analysis the RPI is described by a 
Schr\"odinger equation with a deterministic anti-Hermitian term
representing the effect of the known measurement result.
In the {\it a priori} analysis the same equation becomes a stochastic
nonlinear equation in which the wave function and the measurement
result are evaluated at the same time considering their
mutual influence.

In Ref. \cite{POT} we also introduced the concept of 
non-selective measurements corresponding to a probabilistic 
description of the system during a continuous
measurement independently of the result of the measurement itself.
In particular, the RPI approach for selective measurements was shown 
to be equivalent, in the case of non-selective measurements, to a Lindblad 
semi-group equation for the density matrix of the measured system. 
As in the case of selective measurements, the Lindblad equation
describing non-selective processes contains a phenomenological parameter. 
This last can be arbitrarily varied to represent highly or moderately 
accurate measurements.
The Lindblad equation was used in Section 4 of \cite{POT} to analyze 
the appearance of the quantum Zeno effect in a two-level system.
The corresponding results are in qualitative agreement with
the different selective analysis done in \cite{OPT,TPO} 
and compare rather well for a sufficiently high value of the meter 
accuracy with an actual non-selective measurement, 
namely the atomic spectroscopy data of the landmark experiment 
described in \cite{I}.
  
In conclusion, an unbiased reading
of the papers \cite{OPT,TPO} shows that there is no 
{\em serious methodological error} in them and that the developed 
formalism holds for generic, time-dependent measurements of energy.  
Furthermore, what Mensky considers an {\em entirely new type of 
measurement - monitoring of a quantum transition} \cite{M1} was 
already introduced and discussed in \cite{POT}. 
We hope that this Comment may contribute to reconstruct a fair 
perspective on the issue of continuous energy measurements, 
correcting the misrepresentation made in the review paper \cite{M1}.

\end{multicols}
\end{document}